\documentclass[showpacs,amsmath,amssymb,prb,superscriptaddress,twocolumn]{revtex4}
\usepackage[T1]{fontenc}
\usepackage{amssymb,amsmath}
\usepackage{times}
\usepackage{graphicx}
\usepackage{wasysym}
\usepackage{subfigure}
\usepackage[sort&compress]{natbib}
\usepackage{bm}
\usepackage{overpic}
\usepackage{color}
\usepackage{multirow}

\begin{document}

\title{Comparison of Chiral Metamaterial Designs for Repulsive Casimir Force}

\author{R.~Zhao}
\affiliation{Ames Laboratory and Dept.~of Phys.~and Astronomy,
             Iowa State University, Ames, Iowa 50011, U.S.A.}
\affiliation{Applied Optics Beijing Area Major Laboratory, Department of Physics,
Beijing Normal University, Beijing 100875, China}

\author{Th.~Koschny}
\affiliation{Ames Laboratory and Dept.~of Phys.~and Astronomy,
             Iowa State University, Ames, Iowa 50011, U.S.A.}
\affiliation{Institute of Electronic Structure and Laser, FORTH,
             and Department of Materials Science and Technology, University of Crete,71110 Heraklion, Crete, Greece}

\author{E.~N.~Economou}
\affiliation{Institute of Electronic Structure and Laser, FORTH,
             and Department of Materials Science and Technology, University of Crete,71110 Heraklion, Crete, Greece}

\author{C.~M.~Soukoulis}
\affiliation{Ames Laboratory and Dept.~of Phys.~and Astronomy,
             Iowa State University, Ames, Iowa 50011, U.S.A.}
\affiliation{Institute of Electronic Structure and Laser, FORTH,
             and Department of Materials Science and Technology, University of Crete,71110 Heraklion, Crete, Greece}
\date{\today}


\begin{abstract}
In our previous work [Phys. Rev. Lett. 103, 103602 (2009)], we found that repulsive Casimir forces could be realized by using chiral metamaterials if the chirality is strong enough. In this work, we check four different chiral metamaterial designs (i.e., Twisted-Rosettes, Twisted-Crosswires, Four-U-SRRs, and Conjugate-Swastikas) and find that the designs of Four-U-SRRs and Conjugate-Swastikas are the most promising candidates to realize repulsive Casimir force because of their large chirality and the small ratio of structure length scale to resonance wavelength.    
\end{abstract}


\pacs{42.50.Ct, 78.20.Ek, 12.20.-m}

\maketitle
\textit{Introduction.} -- Two neutral conducting surfaces separated by a vacuum attract each other due to the quantum fluctuations of the vacuum field \cite{Casimir, Lifshitz} and mutual polarization. The force is named \textit{Casimir force} after H. B. G. Casimir who discovered this force by considering the vacuum energy of the electromagnetic field between the plates. \cite{Casimir} Recently, this field is attracting increasing interest theoretically \cite{theory_Yannopapas, theory_Zhao, theory_Lambrecht, theory_Dalvit, theory_Rosa, theory_Rodriguez} and experimentally. \cite{exp_Kim, exp_Man, exp_Masuda, exp_Munday, exp_Chan, exp_Hertlein} The original Casimir force is always monotonic and attractive; however, the polarization mechanism may possibly lead to repulsive Casimir force. To this goal, some researchers immersed the interacting two plates in a fluid;\cite{exp_Munday, repulsive_Dzyaloshinskii} Some used an asymmetric setup of mainly (purely) electric/vacuum/mainly (purely) magnetic;\cite{theory_Yannopapas, theory_Rosa, repulsive_Boyer,repulsive_Kenneth} Some employed a perfect lens sandwiched between the interacting plates;\cite{repulsive_Leonhardt} And some turned to complex geometries.\cite{repulsive_Rodriguez} In our recent previous work,\cite{theory_Zhao} we found repulsive Casimir forces could be realized by using chiral metamaterials (CMMs) if the chirality is strong enough. This work has met considerable interest because it showed that a repulsive Casimir force between two chiral media separated by vacuum could be obtained without the need for magnetic materials. This approach to a repulsive Casimir force could be confirmed numerically \cite{numerical_Rodriguez, numerical_McCauley, numerical_Reid} or experimentally once CMM designs with large enough chirality are obtained. Such metamaterials require strong chiral response and a structural length scale much smaller than their resonance wavelength. Therefore, in this work, we will check four different layered CMM designs (i.e., Twisted-Rosettes, Twisted-Crosswires, Four-U-SRRs, and Conjugate-Swastikas) for their potential to realize repulsive Casimir force. Twisted-Rosettes \cite{rosette_Plum} and Twisted-Crosswires \cite{crosswires_Decker, crosswires_Jiangfeng} designs have been published before; Four-U-SRRs and Conjugate-Swastikas are new designs for chiral metamaterials. 

In the following, the extended Lifshitz theory for calculating Casimir force in CMMs will be briefly introduced first. Then we will derive the analytical form of frequency dependence of the constitutive parameters (permittivity $\epsilon$, permeability $\mu$, and chirality $\kappa$) of CMMs based on effective LC circuit approach, and a retrieval of the response function for CMMs will be given. Then the response functions retrieval from numerical simulation will be used to compare four CMM designs to find the most promising candidate to possibly realize a repulsive Casimir force.

\textit{Extended Lifshitz Theory.} --
Lifshitz \cite{Lifshitz} generalized the calculation of Casimir force between two media characterized by frequency-dependent
dielectric functions $\epsilon_1(\omega)$ and $\epsilon_2(\omega)$.
Subsequently, there was further generalization to general
bi-anisotropic media.\cite{Barash} The formula for the force or
the interaction energy per unit area can be expressed in terms of
the reflection amplitudes, $r_j^{ab}$ ($j=1,2$),\cite{Lambrecht} at the interface between vacuum and medium $j$, giving the ratio of the reflected EM wave of
polarization \textit{a} by the incoming wave of polarization
\textit{b}. Each \textit{a} and \textit{b} stands for either
electric (TM or p) or magnetic (TE or s) waves. The frequency
integration is performed along the imaginary axis by setting
$\omega=i\xi$. The interaction energy per unit
area becomes
\[\frac{E(d)}{A}=\frac{\hbar}{2\pi}\int_0^\infty d\xi\int\frac{d^2\mathbf{k}_\parallel}{(2\pi)^2}\ln\det \mathbf{G};\hspace{3mm}\mathbf{R}_j=\left\lvert\begin{array}{cc}
r_j^\textrm{ss}&r_j^\textrm{sp}\\r_j^\textrm{ps}&r_j^\textrm{pp}
\end{array}\right\rvert,
\]
where $\mathbf{G}=1-\mathbf{R}_1\cdot\mathbf{R}_2e^{-2Kd},~K=\sqrt{\mathbf{k}_\parallel^2+\xi^2/c^2}$. 
Explicit expressions for the elements $r_j^{ab}$ from the vacuum to CMM are available.\cite{Lakhtakia} The diagonal terms are $r_j=[\mp\Gamma_-(\chi_++\chi_-)-(\chi_+\chi_--1)]/\Delta$ for $s$ and $p$ respectively, and $r_j^\textrm{sp}=-r_j^\textrm{ps}=i(\chi_+-\chi_-)/\Delta$; where $\Delta=\Gamma_+(\chi_++\chi_-)+(\chi_+\chi_-+1)$, $\chi_\pm=\sqrt{\mathbf{k}_\parallel^2+n_\pm^2\xi^2/c^2}/n_\pm K$, $\Gamma_\pm=(\eta^2_0\pm\eta^2_j)/2\eta_0\eta_j$, $n_\pm=\sqrt{\epsilon_j\mu_j}\pm \kappa_j$, $\eta_0=\sqrt{\mu_0/\epsilon_0}$,
$\eta_j=\sqrt{\mu_0\mu_j/\epsilon_0\epsilon_j}$.
$\epsilon_j$, $\mu_j$, and $\kappa_j$ are respectively the relative permittivity, the relative permeability, and the chirality coefficients of the medium $j$. $\epsilon_0$ and $\mu_0$ are the permittivity and permeability of the vacuum. 

\textit{Constitutive Parameters of CMMs.} -- To find the appropriate frequency dependencies of the CMM effective parameters, we consider the example of the simplest chiral resonator as show in Fig. \ref{helix}. The structure consists of an open circular
loop and two short wires. The area of the loop is $S=\pi r^2$ and
the length of each wire is $l$. The wires are perpendicular to the
loop and connected to the ends of the open loop. In
a homogeneous external field, the driving electric potential can be
written as:
\begin{figure}[htb!]
 \centering{\includegraphics[angle=0, width=0.4\textwidth]{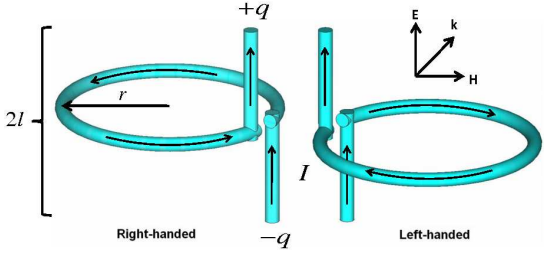}}
 \caption{(Color online) Schematics of the single right-handed (left) and left-handed (right) helix structures.}
 \label{helix}
\end{figure}
\begin{equation}
U=2lE_0\pm\mu_0S\dot{H}_0,
\end{equation}
where $\pm$ signs correspond to right-handed and left-handed helix
resonators. Applying
the effective RLC circuit model, we have
\begin{equation}
L \dot{I}+\frac{q}{c}+RI=U,\ \ \ \ I=\dot{q}.
\end{equation}
Let $\alpha=\frac{l}{L}$, $\beta=\mu_0\frac{A}{L}$, $\gamma=\frac{R}{L}$, and
$\omega^2_0=\frac{1}{LC}$. Assuming the external fields are
time harmonic, $\propto e^{-i\omega t}$, we get the solution: 
\begin{equation}
q=\frac{\alpha}{-\omega^2-i\omega\gamma+\omega_0^2}E_0\pm\frac{i\omega\beta}{-\omega^2-i\omega\gamma+\omega_0^2}H_0.
\end{equation}
Because the electric dipole, $\mathbf{p}=q\mathbf{l}$; the
magnetic dipole, $\mathbf{m}=\pm
I\mathbf{A}=\pm\dot{q}\mathbf{A}$; and $\mathbf{D}=\epsilon_0\mathbf{E}+\mathbf{P}$, $\mathbf{B}=\mu_0\mathbf{H}+\mu_0\mathbf{M}$, we have
\begin{subequations}\label{DEP}
\begin{eqnarray}
\mathbf{D}\!\!\!&=&\!\!\!\epsilon_0\mathbf{E}\!+\!\frac{\alpha l N/V_0}{\omega_0^2-\omega^2-i\omega\gamma}\mathbf{E}\!+\!\frac{\pm i\omega\beta l N/V_0}{\omega_0^2-\omega^2-i\omega\gamma} \mathbf{H},\\
\mathbf{B}\!\!\!&=&\!\!\!\mu_0\mathbf{H}\!+\!\frac{\mp
i\mu_0 \omega\alpha A N/V_0}{\omega_0^2-\omega^2-i\omega\gamma}\mathbf{E}\!+\!\frac{\omega^2\mu_0\beta A N/V_0}{\omega_0^2-\omega^2-i\omega\gamma}\mathbf{H}.
\end{eqnarray}
\end{subequations}
Here, the directions of $\mathbf{l}$ and $\mathbf{A}$ are merged into $E_0$ and $H_0$ to form the vectors of $\mathbf{E}$ and $\mathbf{H}$. $V_0$ and $N$ denotes the volume of one unit cell and the number of resonators in one unit cell. Therefore, the constitutive equation can be written as: 
\begin{equation}\label{constitutive}
\left(\begin{array}{cc}\mathbf{D}\\\mathbf{B}\end{array}\right)=\left(\begin{array}{cc}
\epsilon_0\epsilon&i\kappa/c_0\\-i\kappa/c_0&\mu_0\mu
\end{array}\right)\left(\begin{array}{cc}\mathbf{E}\\\mathbf{H}\end{array}\right),
\end{equation}
and the constitutive parameters have the following forms of frequency dependent: 
\begin{subequations}\label{epsilon_mu_kappa}
\begin{eqnarray}
\epsilon&=&1+\frac{\Omega_\epsilon \omega_0^2}{\omega_0^2-\omega^2-i\omega\gamma},\\
\mu&=&1+\Omega_\mu+\frac{\Omega_\mu \omega^2}{\omega_0^2-\omega^2-i\omega\gamma},\\
\kappa&=&\frac{\pm \Omega_\kappa \omega_0\omega}{\omega_0^2-\omega^2-i\omega\gamma},
\end{eqnarray}
\end{subequations}
where $\Omega_\epsilon$, $\Omega_\mu$, and $\Omega_\kappa$ are the coefficients of the resonance terms in $\epsilon$, $\mu$, and $\kappa$, i.e.  $\Omega_\epsilon=\frac{\alpha l N}{ V_0 \epsilon_0 \omega_0^2}$, $\Omega_\mu=\frac{\beta A N}{V_0}$, and $\Omega_\kappa=\frac{\beta l c_0 N}{V_0 \omega_0}=\frac{\mu_0 c_0 \alpha A N}{V_0 \omega_0}$, which describe the strength of the resonance. The $\Omega_\mu$ in the constant term of $\mu$ is introduced to make sure the physically correct limiting behavior, $\mu(\infty)=1$, is obeyed. Eqs. (\ref{epsilon_mu_kappa}) will be used to calculate the Casimir force directly, therefore, the Casimir force is determined by five parameters, $\Omega_\epsilon$, $\Omega_\mu$, $\Omega_\kappa$, $\omega_0$, and $\gamma$. But when using these equations to fit the numerical simulation results, $\epsilon$ and $\mu$ need some adjustments, i.e., changing the constant term, $1$, to $\epsilon_b$ in $\epsilon$ and changing the constant term, $1+\Omega_\mu$, to $\mu_b$ in $\mu$ . This is reasonable, because $\epsilon_b$ and $\mu_b$ depend on the properties of the bound electron in each material. 

\textit{Parameter Retrieval for CMMs.} --
Parameter retrieval \cite{retrievalref_Smith} is a basic technique to obtain the electromagnetic properties of the effective media. The effective media are usually considered to be homogeneous when the size of the unit cell of the structure is much smaller than the wavelength $\lambda$. The constitutive parameters, $\epsilon$ and $\mu$ or $n=\sqrt{\epsilon\mu}$ and $z=\sqrt{\mu/\epsilon}$,  are well defined and can be determined from reflection and transmission coefficients ($S$ parameters). For CMMs, the impedance and refractive index for right (left) circularly polarized fields, $n_\pm$ (+ $\leftrightarrow$ RCP, - $\leftrightarrow$ LCP), can be expressed as:\cite{rosette_Plum} 
\begin{subequations}\label{impedance}
\begin{align}
&z=\pm\sqrt{\frac{(1+R)^2-T_+T_-}{(1-R)^2-T_+T_-}},\\
&n_\pm=\frac{i}{k_0d}\{\ln[\frac{1}{T_\pm}(1-\frac{z-1}{z+1}R)]\pm2m\pi\},
\end{align}
\end{subequations}
where $R$ is reflection coefficient (the reflections, $R_+$ and $R_-$, of RCP and LCP waves are the same), $T_+$ and $T_-$ are the transmission coefficients of RCP and LCP respectively, $m$ is an integer determined by the proper choice of the branch. The sign determination and the proper choice of the branch in Eqs. (\ref{impedance}) are based on the conditions:
\begin{equation}\label{restrict}
\Re(z)\geq0, \Im(n)\geq0,
\end{equation}
which are needed by the energy conservation principle. Then the other parameters can be obtained as: $n=(n_++n_-)/2,~\kappa=(n_+-n_-)/2,~\epsilon=n/z$, and $\mu=nz$.

\textit{Repulsive force for four different CMMs.}--
The above helix structure can give us very ideal chirality, but this three-dimensional structure is impossible to be fabricated experimentally. It's better to go to some layered-like structures called layered CMMs, which can be fabricated at optical frequencies by using a layer-by-layer technique.\cite{LiuNa} In the following, we will check two published chiral designs, Twisted-Rosettes \cite{rosette_Plum} and Twisted-Crosswires,\cite{crosswires_Decker, crosswires_Jiangfeng} and two new designs, Four-U-SRRs and Conjugate-Swastikas. 

\begin{table*}
\begin{tabular}{ | c | c | c | c |p{4.0cm} | }
  \hline
    \subfigure{\includegraphics[width=0.12\textwidth]{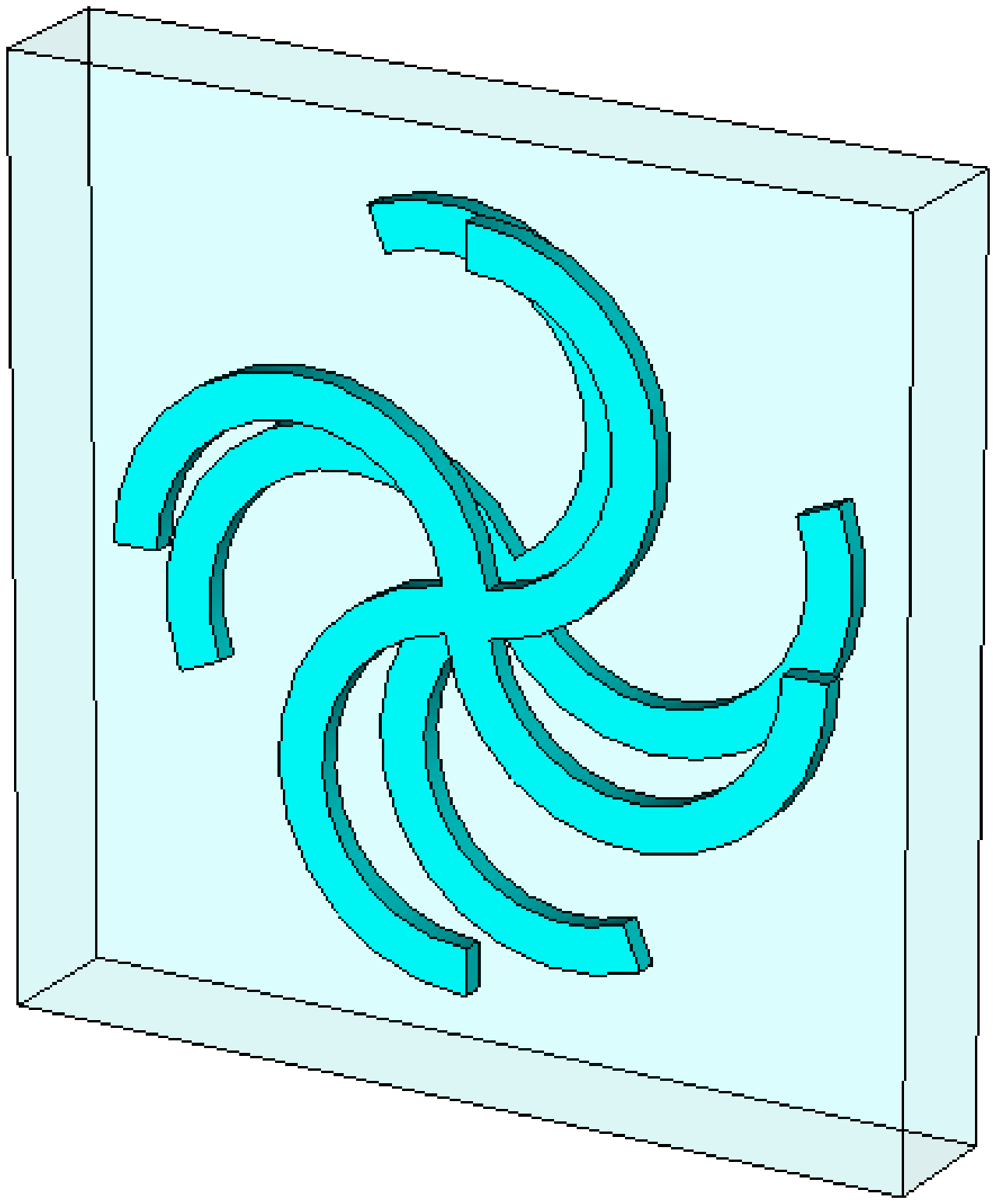}}
 & \subfigure{\includegraphics[width=0.20\textwidth]{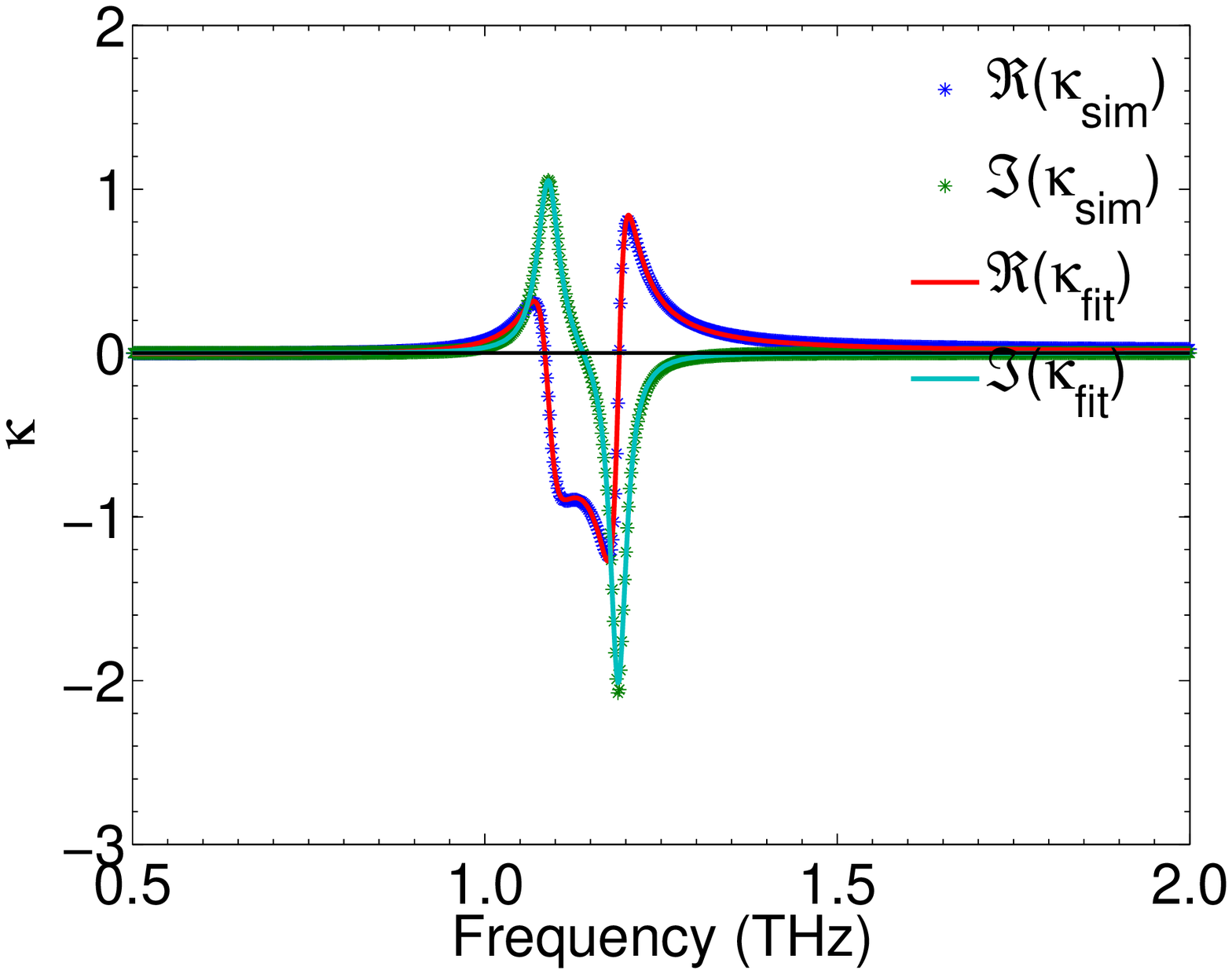}}
 & \subfigure{\includegraphics[width=0.20\textwidth]{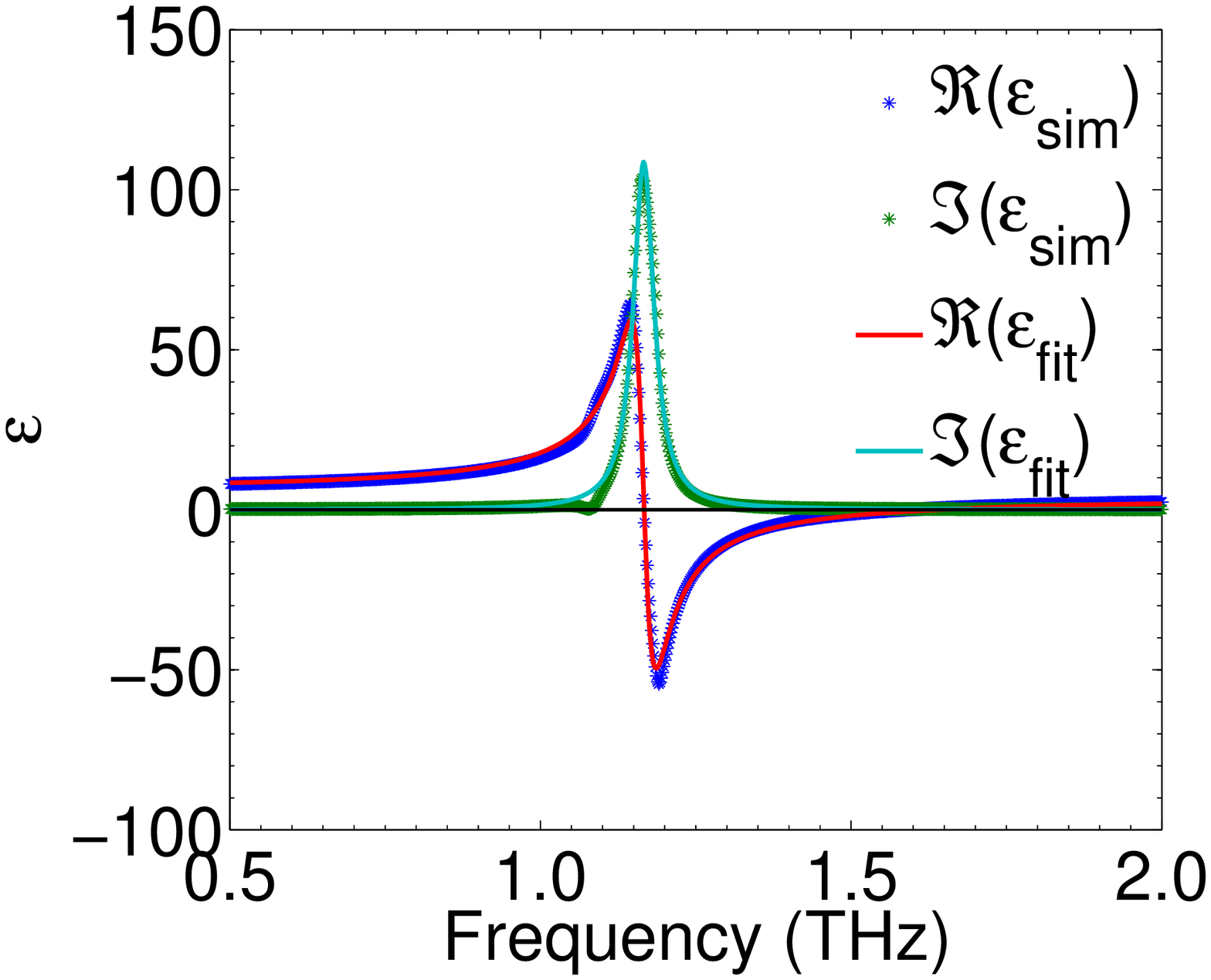}}
&\subfigure{\includegraphics[width=0.20\textwidth]{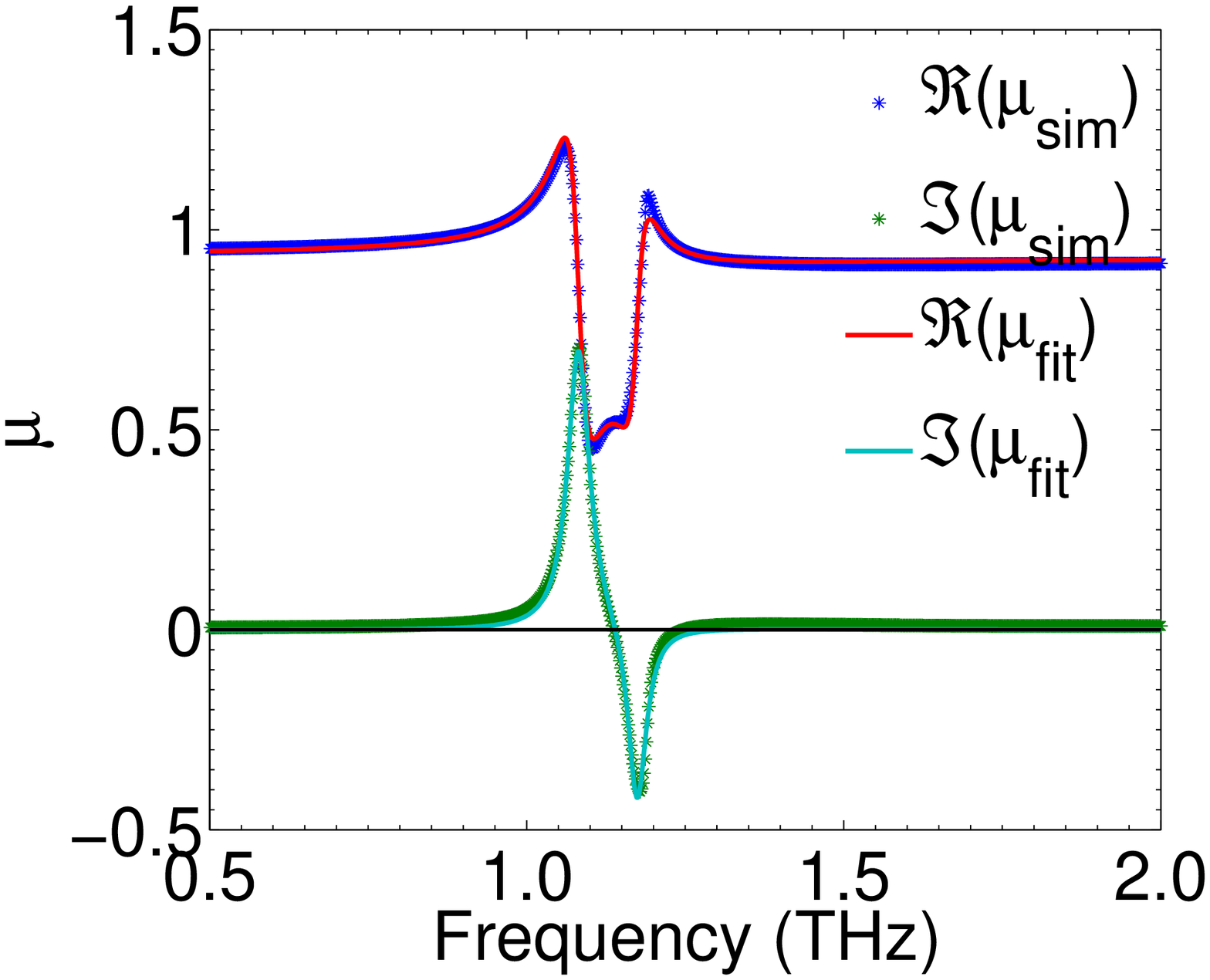}}
&$\epsilon_b=3.8418$, $\mu_b=0.9423$,
 $\Omega_{\epsilon 1}=None$, $\Omega_{\epsilon 2}=3.7514$,
$\Omega_{\mu 1}=0.0279$, $\Omega_{\mu 2}=0.0141$,
$\Omega_{\kappa 1}=0.0384$, $\Omega_{\kappa 2}=0.0490$,
$\omega_{10}=1.0857$, $\omega_{20}=1.1765$,
$\gamma_1=0.0373$, $\gamma_2=0.0296$.
(Medium size and small chirality)\\
  \hline 
     \subfigure{\includegraphics[width=0.12\textwidth]{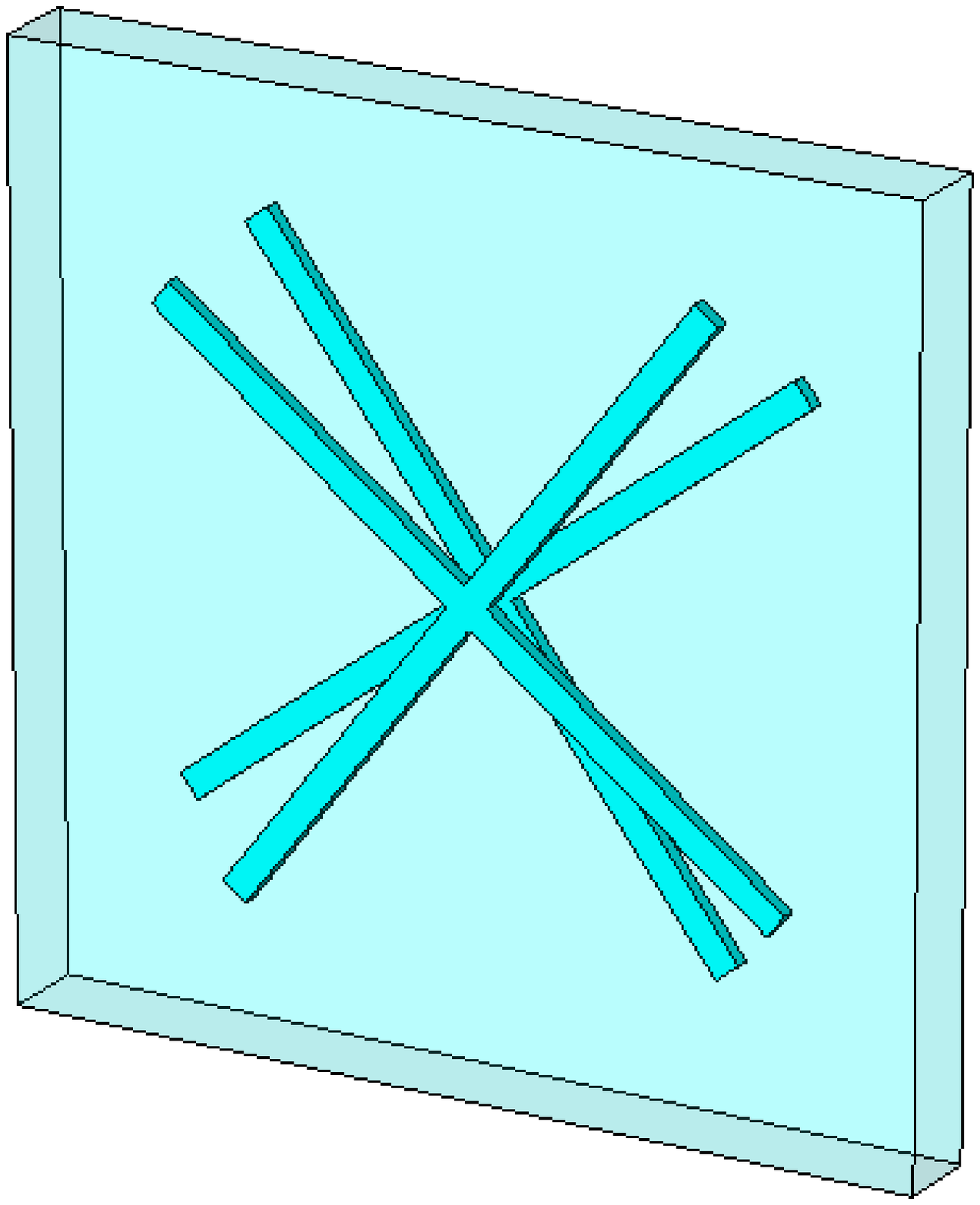}}
 & \subfigure{\includegraphics[width=0.20\textwidth]{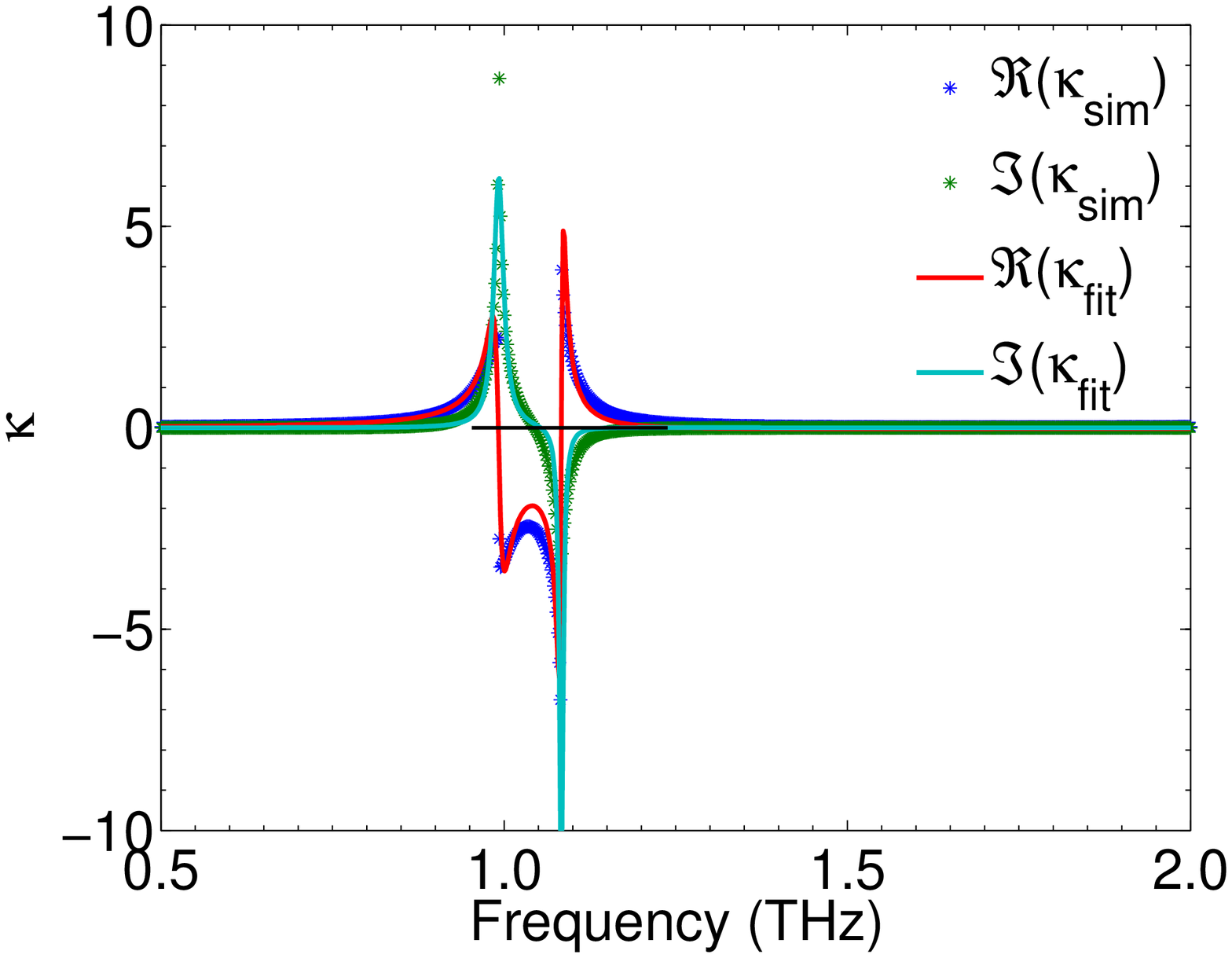}}
 & \subfigure{\includegraphics[width=0.20\textwidth]{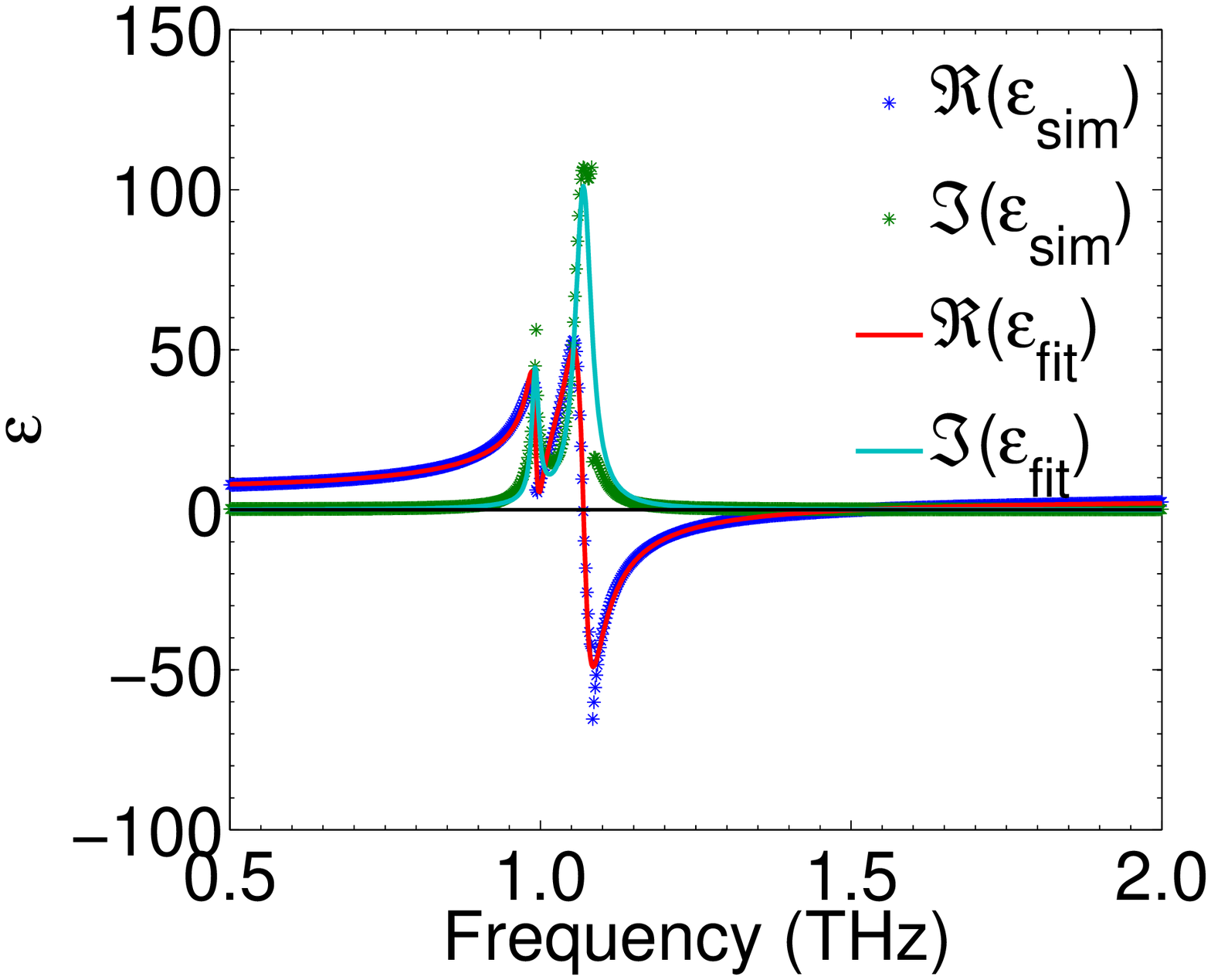}}
&\subfigure{\includegraphics[width=0.20\textwidth]{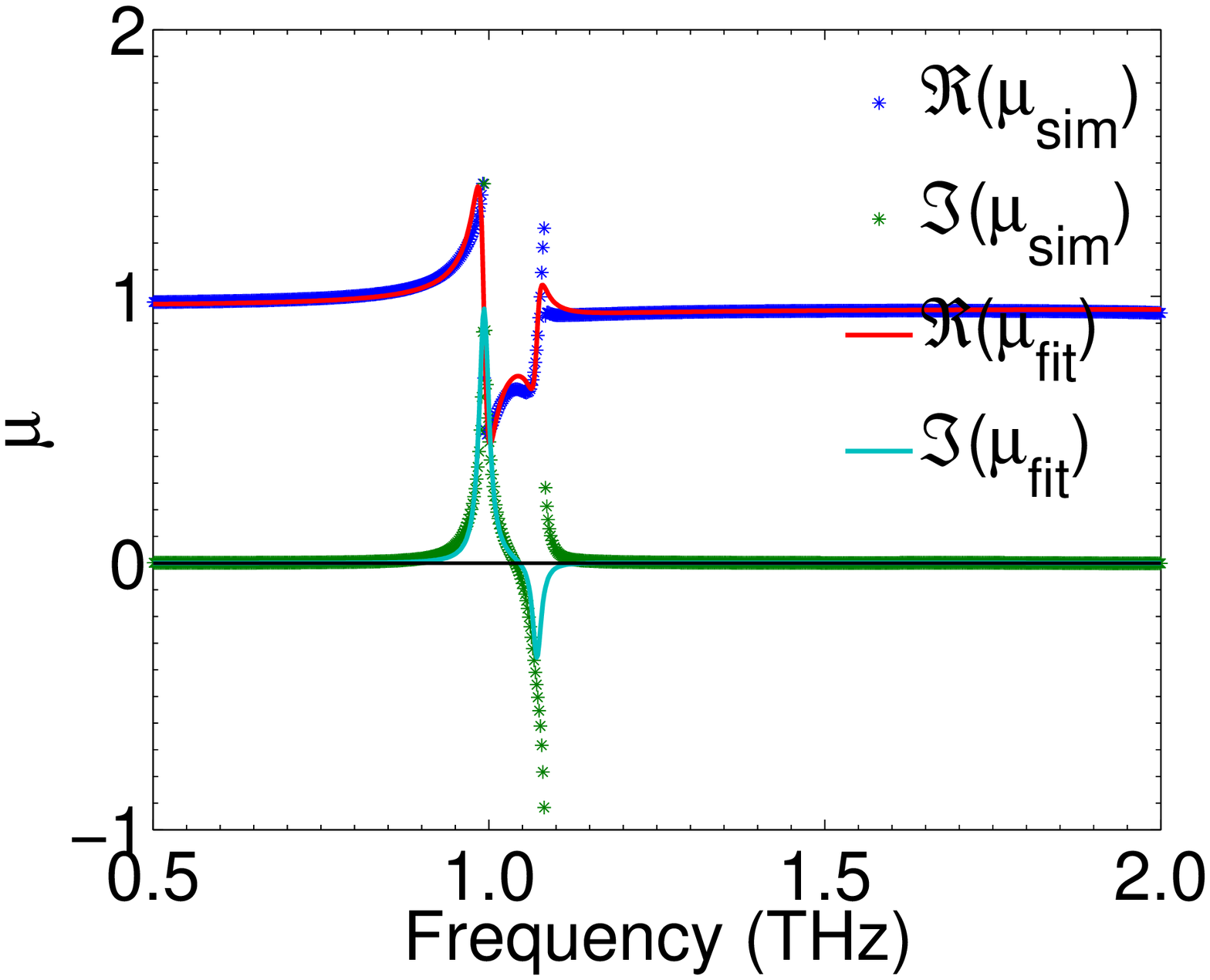}}
&
$\epsilon_b=3.3422$, $\mu_b=0.9696$,
$\Omega_{\epsilon 1}=0.6232$, $\Omega_{\epsilon 2}=2.8806$,
$\Omega_{\mu 1}=0.0177$, $\Omega_{\mu 2}=0.0046$,
$\Omega_{\kappa 1}=0.1007$, $\Omega_{\kappa 2}=0.0961$,
$\omega_{10}=0.9930$, $\omega_{20}=1.070$,
$\gamma_1=0.0175$, $\gamma_2=0.0175$.
(Large size and $\Omega_{\kappa_1}/\omega_\kappa^c=25.1\%$.)

\\
  \hline
  \subfigure{\includegraphics[width=0.12\textwidth]{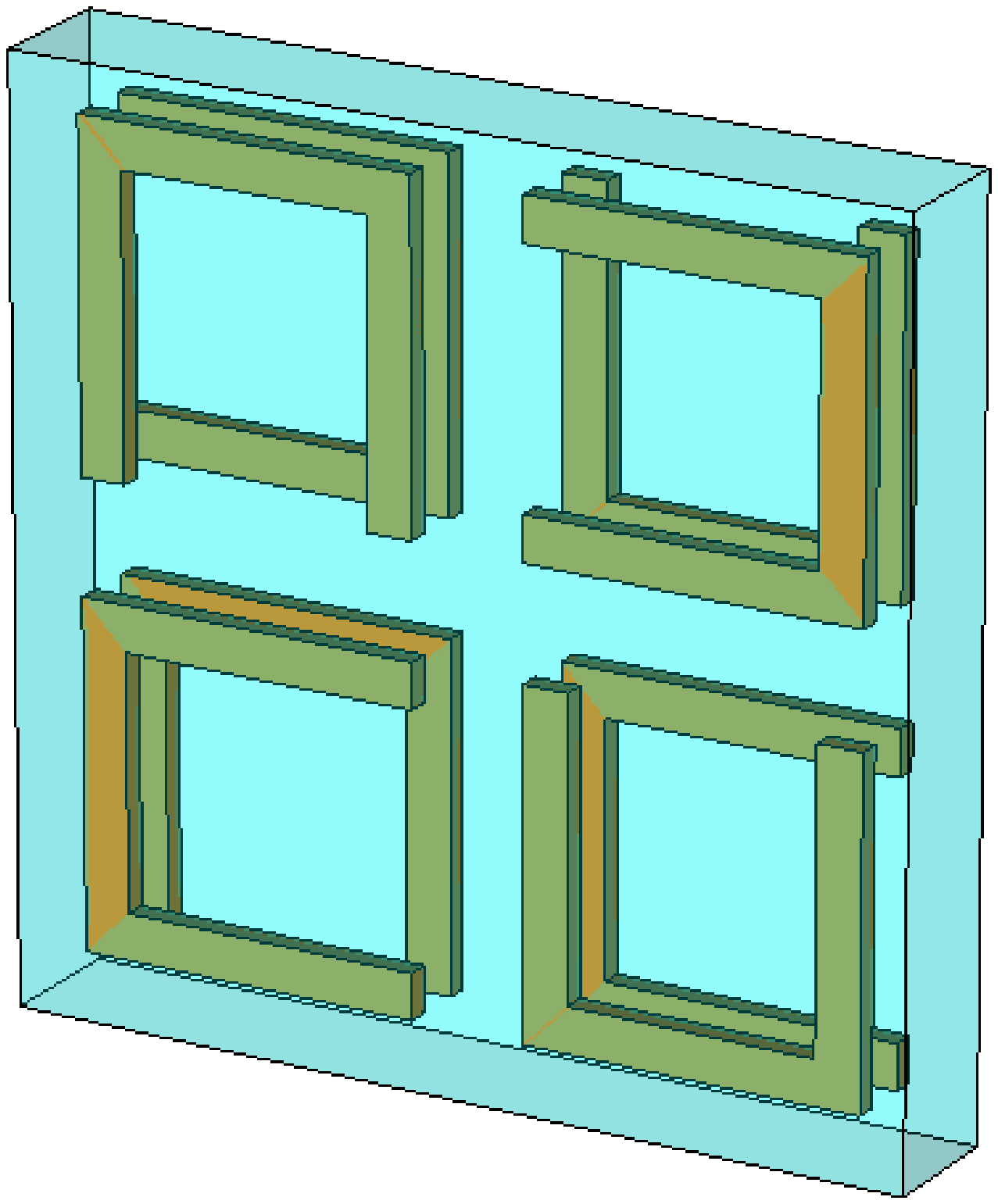}}
 & \subfigure{\includegraphics[width=0.20\textwidth]{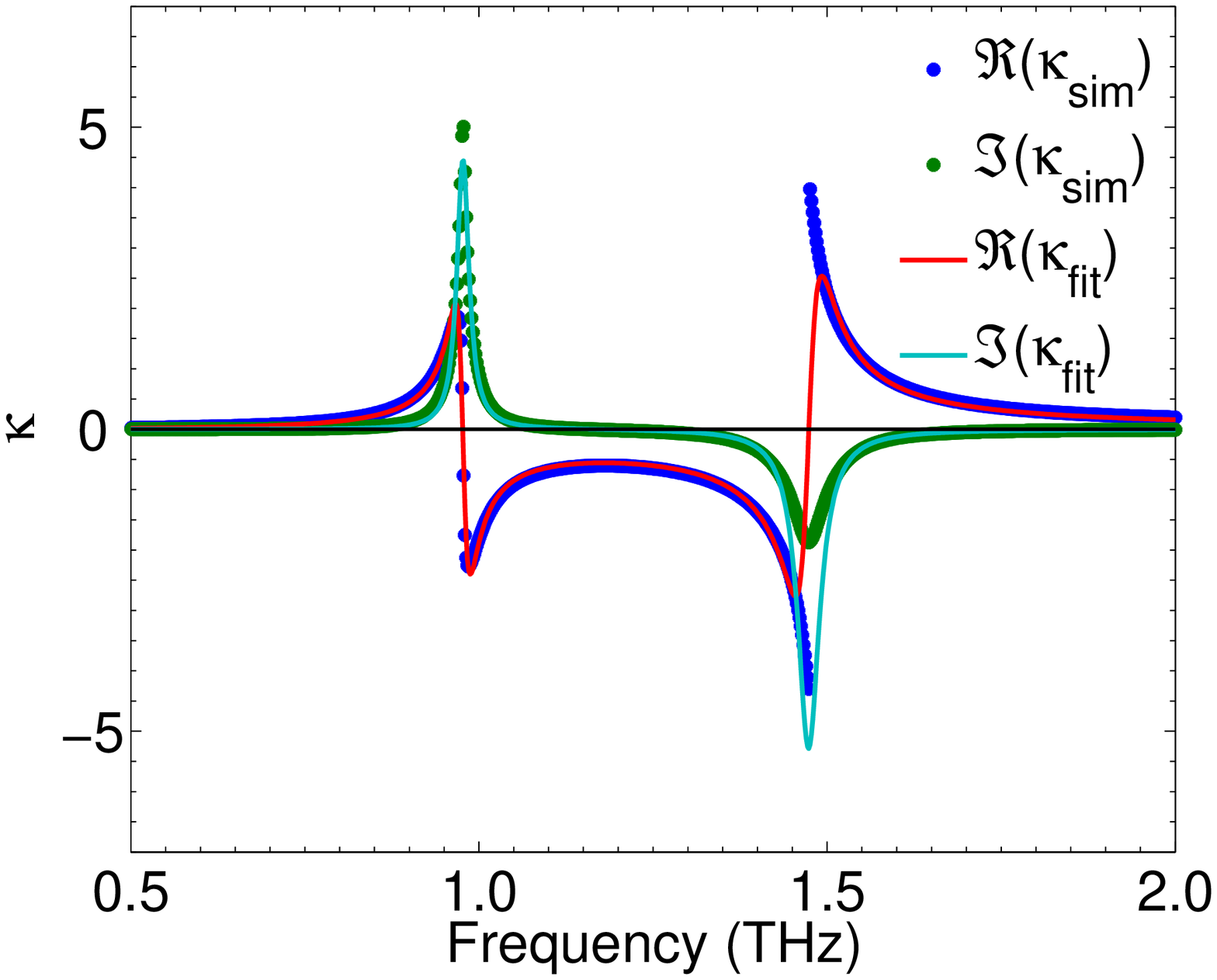}}
 & \subfigure{\includegraphics[width=0.20\textwidth]{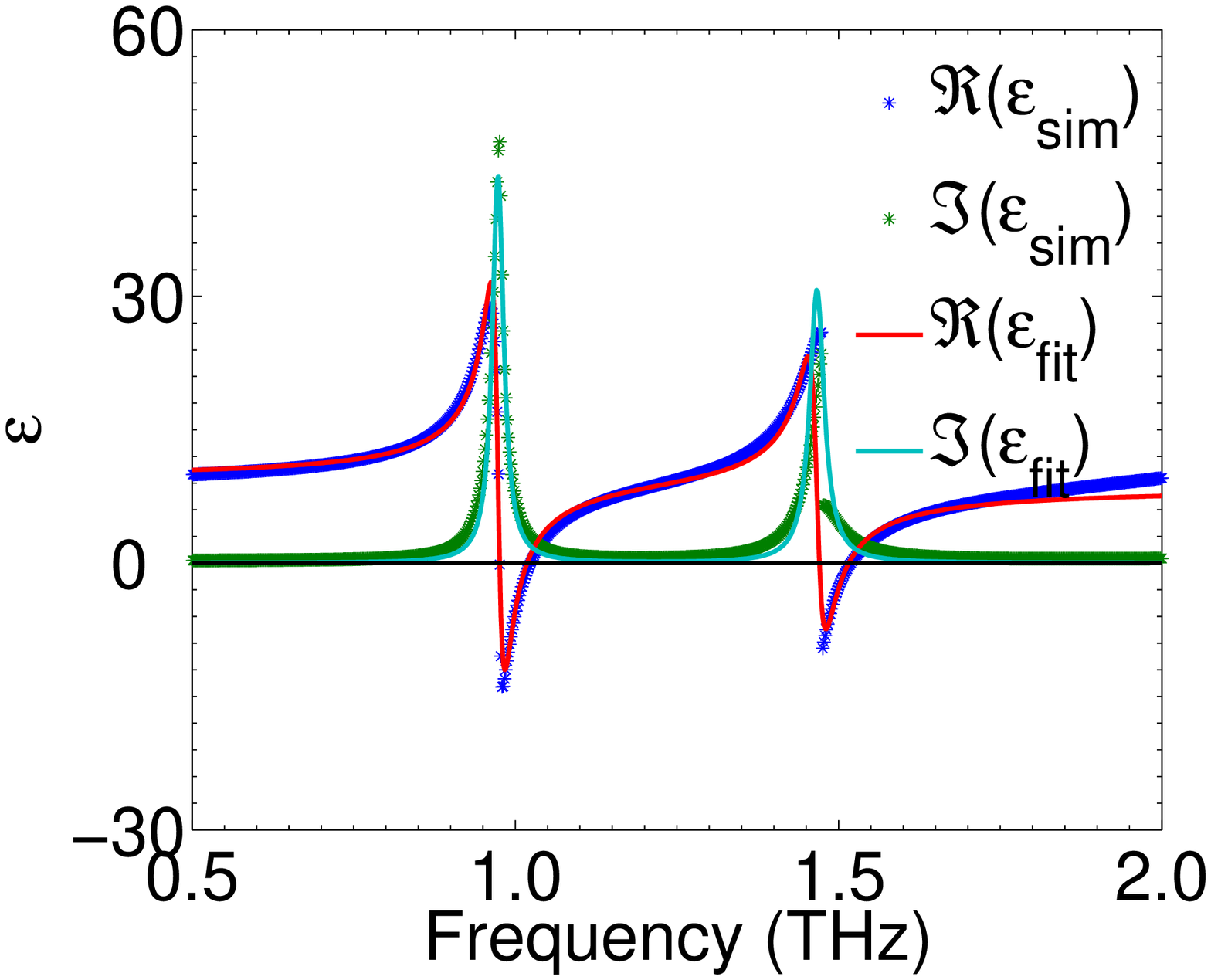}}
&\subfigure{\includegraphics[width=0.20\textwidth]{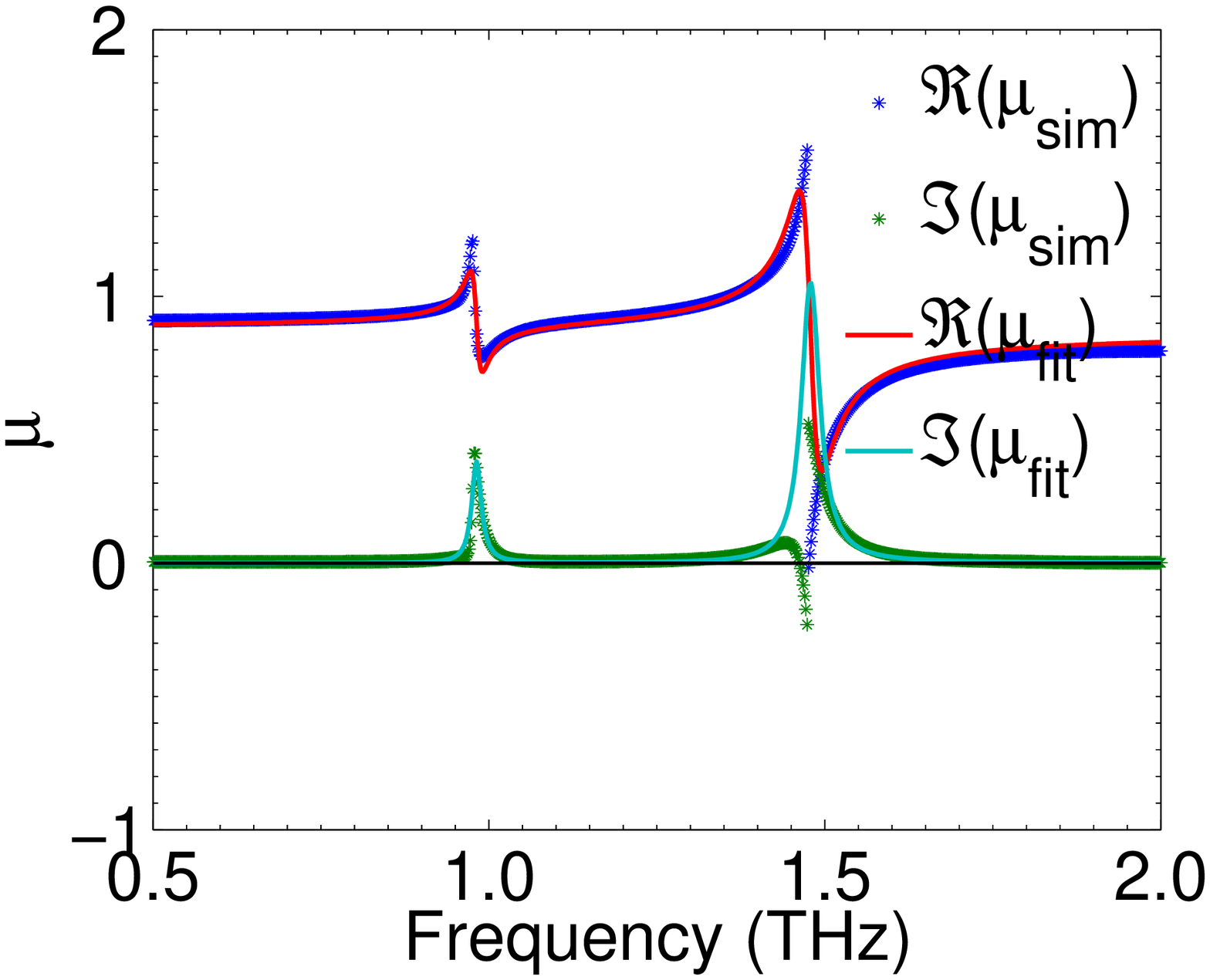}}
&$\epsilon_b=8.5172$, $\mu_b=0.8895$,
$\Omega_{\epsilon 1}=0.9626$, $\Omega_{\epsilon 2}=0.5770$,
$\Omega_{\mu 1}=0.0067$, $\Omega_{\mu 2}=0.0237$,
$\Omega_{\kappa 1}=0.0968$, $\Omega_{\kappa 2}=0.1362$,
$\omega_{10}=0.9773$, $\omega_{20}=1.4729$,
$\gamma_1=0.0204$, $\gamma_2=0.0223$.
(Medium size and $\Omega_{\kappa_1}/\omega_\kappa^c=16.5\%$.)\\
  \hline
  \subfigure{\includegraphics[width=0.12\textwidth]{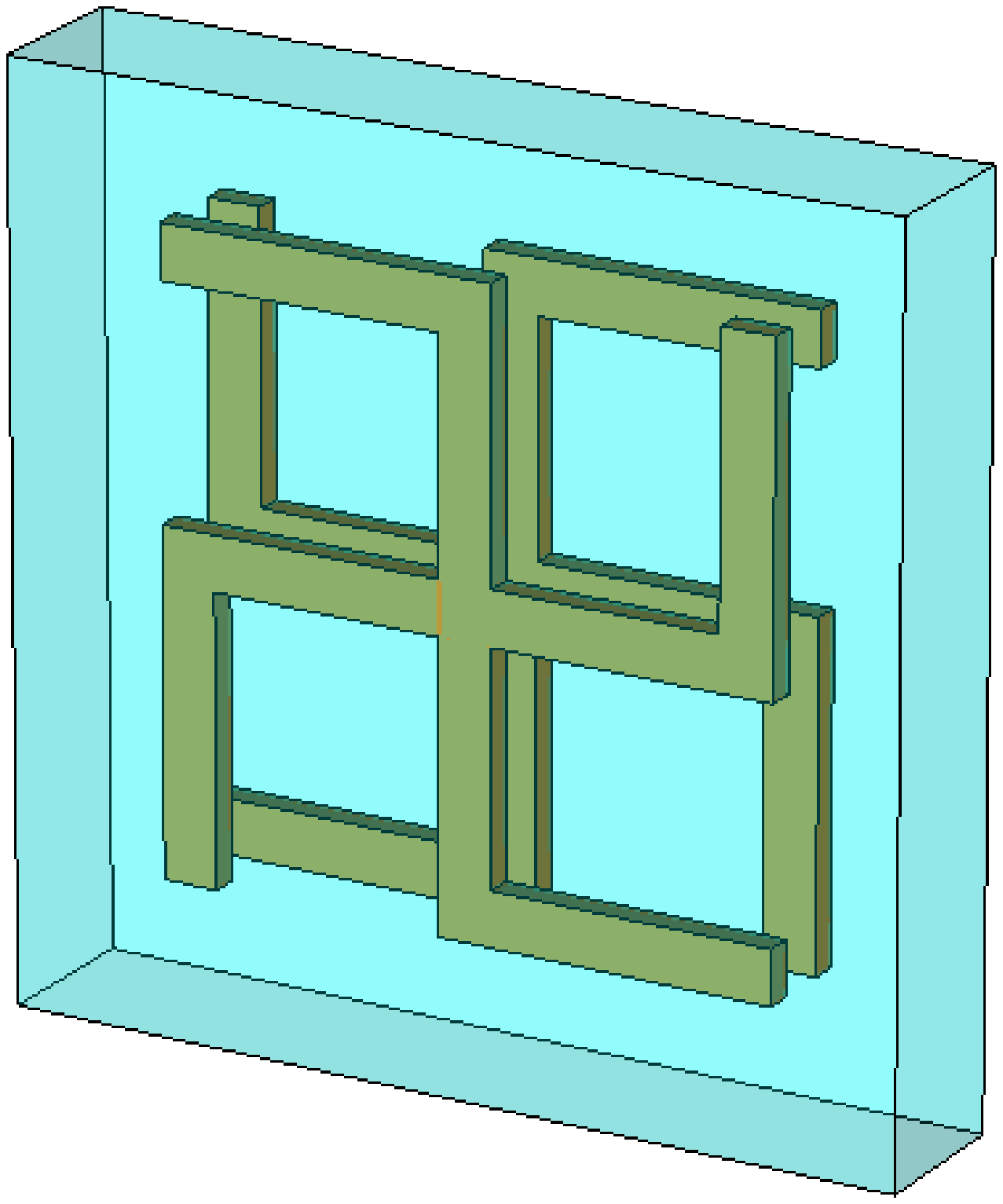}}
 & \subfigure{\includegraphics[width=0.20\textwidth]{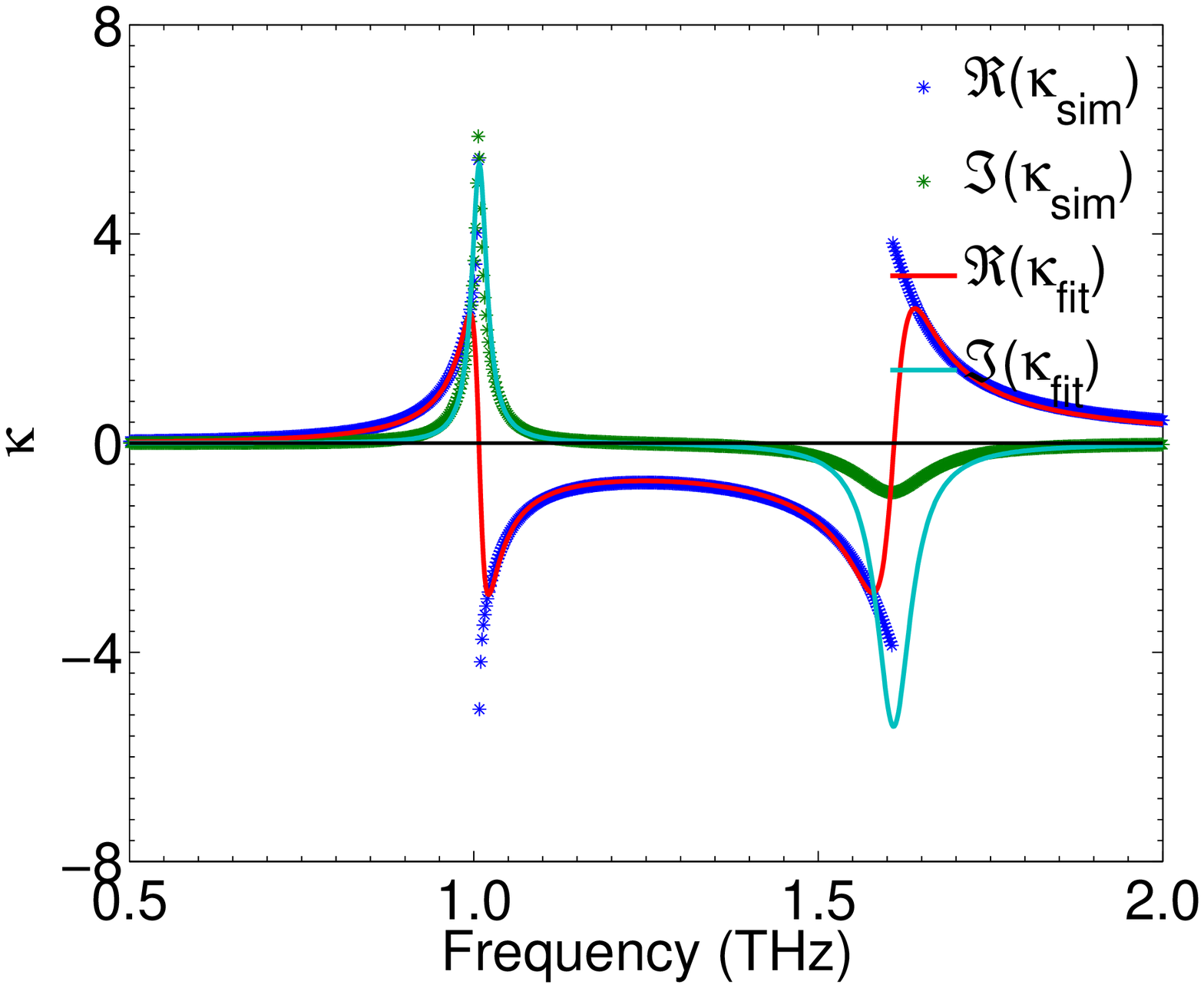}}
 & \subfigure{\includegraphics[width=0.20\textwidth]{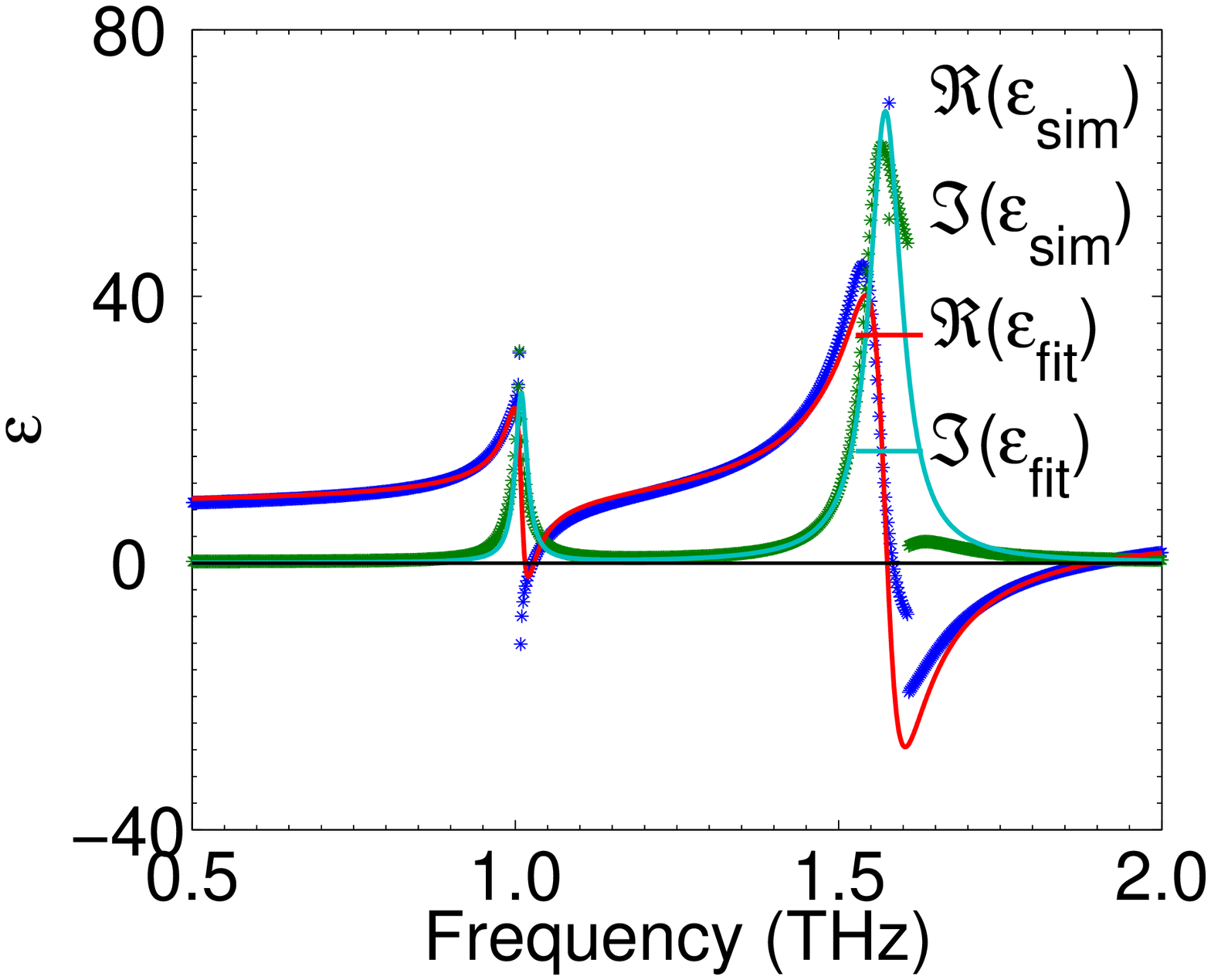}}
&\subfigure{\includegraphics[width=0.20\textwidth]{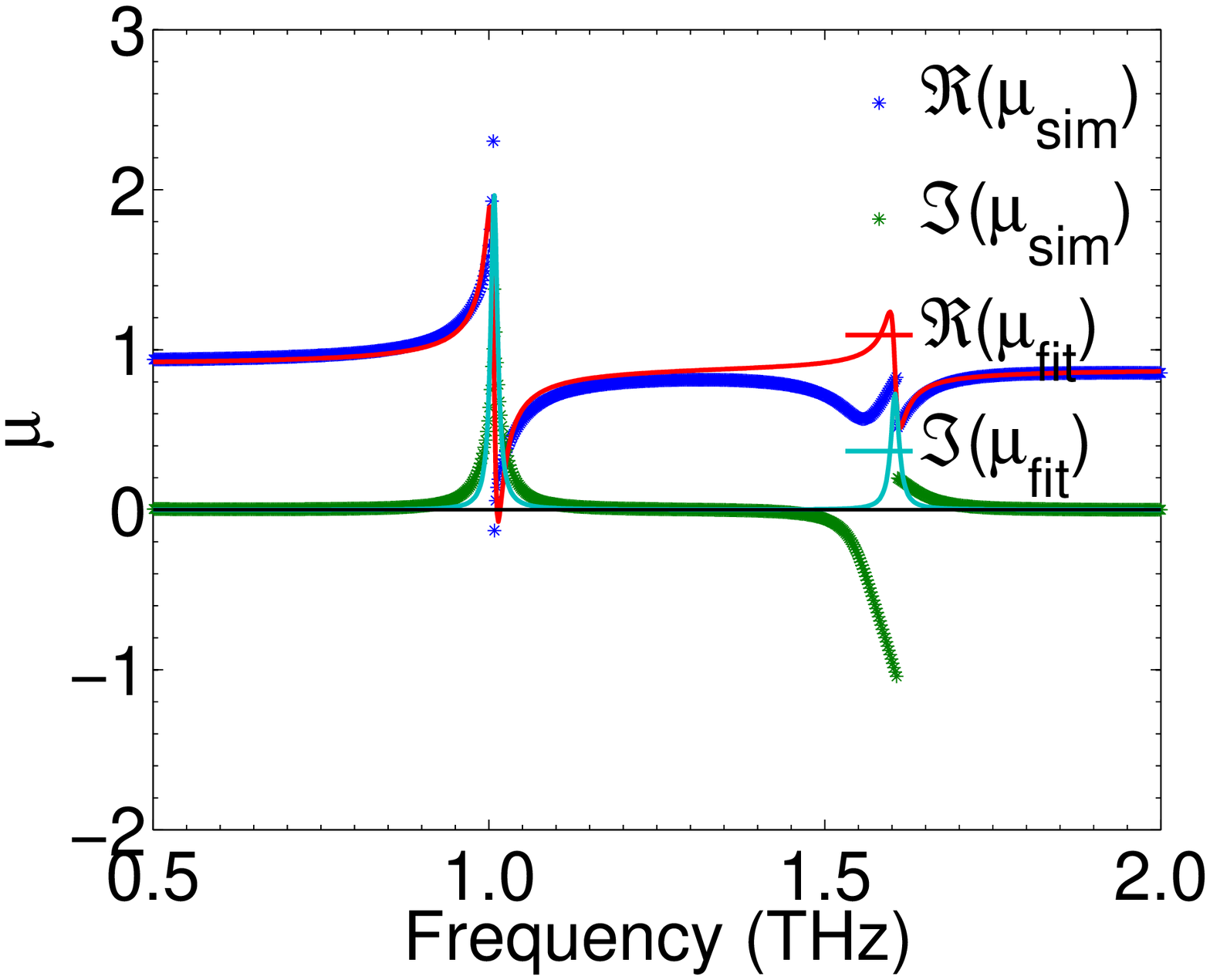}}
&$\epsilon_b=6.0111$, $\mu_b=0.9158$,
$\Omega_{\epsilon 1}=0.5497$, $\Omega_{\epsilon 2}=2.6385$,
$\Omega_{\mu 1}=0.0237$, $\Omega_{\mu 2}=0.0067$,
$\Omega_{\kappa 1}=0.1408$, $\Omega_{\kappa 2}=0.2053$,
$\omega_{10}=1.0082$, $\omega_{20}=1.5956$,
$\gamma_1=0.0200$, $\gamma_2=0.0384$.
(Small size and $\Omega_{\kappa_1}/\omega_\kappa^c=39.4\%$.)\\
  \hline
\end{tabular}
  \caption{(Color online) Four kinds of layered CMM designs. The first column shows the structures of the four designs, i.e., Twisted-Rosettes, Twisted-Crosswires, Four-U-SRRs, and Conjugate-Swastikas respectively from the top to the bottom. From second to fourth columns show the simulation and retrieval results (stars/blue and dark-green lines) and fitting results (solid/red and dark-cyan lines) using analytical Eqs. (\ref{epsilon_mu_kappa}). The fifth column shows the resulting fitting parameters. The subscripts of $1$ and $2$  in the fifth column denote the first and the second resonances. $\epsilon_b$ and $\mu_b$ are the background permittivity and permeability. The parenthesis shows the main features of the design.}
  \label{layered chiral metamaterials}
\end{table*}

The first column of Table \ref{layered chiral metamaterials} shows the four kinds of designs of layered CMMs. From the top to the bottom are  Twisted-Rosettes, Twisted-Crosswires, Four-U-SRRs, and Conjugate-Swastikas respectively.  All of the layered CMMs are designed to work at around $1THz$. The metal structure is silver standing in the background of polyimide with $n=2.5$ and the loss tangent $\delta=0.03$. The thickness of the slab of polyimide is $12\mu m$. The distance between the centers of two silver layers is $6\mu m$. The thickness and the width of silver are all $2\mu m$ and $4\mu m$ respectively. In order to get the same frequency, the length of silver are different: For the Twisted-Rosettes, each arm is a semicircle with the radius $r=15\mu m$. The twist angle is $22.5^\circ$. And the size of the unit cell is $V_{UC}=80\mu m\times80\mu m\times12\mu m$. For the Twisted-Crosswires, the length of the wires is $108\mu m$. The twist angle is $15^\circ$. The size of the unit cell is $V_{UC}=126\mu m\times126\mu m\times12\mu m$. For the Four-U-SRRs, the side length of the SRR is $30\mu m$. The size of the unit cell is $V_{UC}=80\mu m\times80\mu m\times12\mu m$. For the Conjugate-Swastikas, the length of the center arm is $49.8\mu m$. The size of the unit cell is $V_{UC}=66.4\mu m\times66.4\mu m\times12\mu m$. 

Using the above retrieval process, Eqs. (\ref{impedance}) and (\ref{restrict}), the constitutive parameters can be obtained from the simulation \cite{CST} results of transmission and reflection for any CMM design. From second to fourth columns of Table \ref{layered chiral metamaterials} show the simulation and retrieval results (stars/blue and dark-green) and fitting results (solid/red and dark-cyan) using adjusted analytical Eqs. (\ref{epsilon_mu_kappa}), i.e., adjust the constant $1$ as $\epsilon_b$ in $\epsilon$ and the constant $1+\Omega_\mu$ as $\mu_b$ in $\mu$ . The fifth column shows the resulting fitting parameters. All of these layered CMMs have two resonances. In the results in Table \ref{layered chiral metamaterials}, some curves are discontinuous at some points at the second resonance. That is because the refraction index $n$ arrives at the edge of the Brillouin edge where the effective medium theory is not valid any more; the metamaterial behaves as a photonic crystal.\cite{photonic crystal_1, photonic crystal_2} The effective parameters are not well defined there. The purpose of including the discontinuous range is to make the fitting to the first resonance much more precise. Comparing the first resonances of these four designs, the chirality of Twisted-Rosettes is the smallest with $\Omega_\kappa=0.0384$. For the Twisted-Crosswires and the newly designed Four-U-SRRs and Conjugate-Swastikas, they all possess large chirality. But for the Twisted-Crosswires, the unit cell size is very large, 3.6 times of the Conjugate-Swastikas. Therefore, the newly designed Four-U-SRRs and Conjugate-Swastikas are better designs with large chirality and small unit cell size. From the view of the complexity of the structures, the Four-U-SRRs is simpler and easier to be fabricated as block CMMs. But from the view of the performance, the Conjugate-Swastikas is better because of larger chirality, $\Omega_\kappa^{Conjugate-Swastikas}$/$\Omega_\kappa^{Four-U-SRRs}$=1.45, and smaller size, $V_{UC}^{Conjugate-Swastikas}/V_{UC}^{Four-U-SRRs}=0.6889$. The lateral length of the Conjugate-Swastikas is $0.22\lambda$, which is a very promising approach to construct a three-dimensional isotropic structure. 

After extracting the strengths of each resonance, $\Omega_\epsilon$, $\Omega_\mu$, and $\Omega_\kappa$, we can use the fitted $\Omega_\epsilon$ and $\Omega_\mu$ to get the critical value of chirality $\omega_\kappa^c$ as we did in our previous paper \cite{theory_Zhao} and then compare with the fitted chirality strength $\Omega_\kappa$ to see how much the percentage we have realized. The critical value of chirality $\omega_\kappa^c$ means that if $\Omega_\kappa>\omega_\kappa^c$, we can obtain the repulsive Casimir force when the separated distance between the interacting plated is smaller than a certain value, otherwise, the Casimir force is always attractive at any distance. For the Four-U-SRRs design, taking the first resonance at $1THz$ as an example, first insert $\Omega_\epsilon=0.9626$, $\Omega_\mu=0.0067$, $\omega_0=0.9773THz$, and $\gamma=0.0204\omega_0$ into Eqs. (\ref{epsilon_mu_kappa}) and then get the critical value $\omega_\kappa^c=0.5856$. Our current Four-U-SRRs design gives us $\Omega_\kappa=0.0968$,  i.e., we have realized 16.5\% ($=\Omega_\kappa / \omega_\kappa^c=0.0968/0.5856$) of critical value. For the Conjugate-Swastikas design with $\Omega_\epsilon=0.5497$, $\Omega_\mu=0.0237$, $\omega_0=1.0082THz$, and $\gamma=0.0200\omega_0$, the critical value is $\omega_\kappa^c=0.3578$. Our current Four-U-SRRs design gives us $\Omega_\kappa=0.1408$, i.e., we have realized 39.4\% ($=\Omega_\kappa / \omega_\kappa^c=0.1408/0.3578$) of critical value. These two designs can be optimized further. As given in our previous paper,\cite{theory_Zhao} the optimized case is $\Omega_\epsilon=\epsilon_\mu$. For the Four-U-SRRs and Conjugate-Swastikas designs, $\Omega_\epsilon\gg \Omega_\mu$, but this is not a general property. $\Omega_\mu$ can possibly be optimized to be close to $\Omega_\epsilon$ by increasing the inductance of the structures, e.g., increasing the space between the two metal layers or decreasing the opening of the SRR.

\textit{Conclusion.}-- In summary, we have checked four different chiral metamaterial designs (i.e., Twisted-Rosettes, Twisted-Crosswires, Four-U-SRRs, and Conjugate-Swastikas) and found out that the designs of Four-U-SRRs and Conjugate-Swastikas are the most promising candidate to realize repulsive Casimir force because of the larger chirality and the smaller ratio of the scale size to the wavelength. However, for the current designs, the critical value for the chiral response has not been reached yet. Between these two designs, Conjugate-Swastikas design is better than Four-U-SRRs. In the future, we will try to optimize the Conjugate-Swastikas designs to reach the critical value of chirality.


Work at Ames Laboratory was supported by the Department of Energy
(Basic Energy Sciences) under contract No.~DE-AC02-07CH11358. This
work was partially supported by the European Community FET project
PHOME (contract No.~213390), US Department of Commerce NIST
70NANB7H6138 and the US Air Force grants. The author Rongkuo Zhao
specially acknowledges the China Scholarship Council (CSC).

\bibliographystyle{apsrev}

\end{document}